\documentclass[conference]{IEEEtran}
\usepackage{blindtext, graphicx}
\usepackage{listings}
\usepackage{float}
\usepackage{graphicx}
\usepackage[font=small]{caption}
 \usepackage{amsmath}
\usepackage[utf8]{inputenc}
\usepackage{multirow}
\usepackage{multicol}
\usepackage{dblfloatfix}

\title{The Derivation and Reconstruction of the Gamma Variate Function for Tracer Dilution Curves}

\author{
   \IEEEauthorblockN{Ishmael N. Amartey\IEEEauthorrefmark{1}, Andreas A. Linninger\IEEEauthorrefmark{2}, Thomas Ventimiglia\IEEEauthorrefmark{3}\IEEEauthorrefmark{4}}
    \IEEEauthorblockA{\IEEEauthorrefmark{1} Department of Statistics and Actuarial Science, Northern Illinois University, Dekalb, Illinois, USA
    \ }
    \IEEEauthorblockA{\IEEEauthorrefmark{2}Department of Biomedical Engineering, University of Illinois at Chicago, Chicago, Illinois, USA
    \ }
    \IEEEauthorblockA{\IEEEauthorrefmark{4} Department of Mathematical Sciences, Northern Illinois University, Dekalb, Illinois, USA
    \\\ }}
\begin{document}

\maketitle
\thispagestyle{empty}
\pagestyle{empty}

%%%%%%%%%%%%%%%%%%%%%%%%%%%%%%%%%%%%%%%%%%%%%%%%%%%%%%%%%%%%%%%%%%%%%%%%%%%%%%%%
\begin{abstract}

Cerebral blood flow and perfusion can be estimated using tracer dilution experiments. Accurate estimation of blood flow parameters is a crucial part of medical imaging for effective diagnosis and treatment. This study explores two themes: (i) the derivation of the gamma variate function as a response tracer infusion and (ii) the estimation of impulse and residue functions from tracer dilution curves parameters via the least squares method.

\end{abstract}

\textbf{\emph{Index Terms}: Tracer dilution, gamma variate curve, parameter estimation, signal reconstruction, least squares method}.

%\end{abstract}

%%%%%%%%%%%%%%%%%%%%%%%%%%%%%%%%%%%%%%%%%%%%%%%%%%%%%%%%%%%%%%%%%%%%%%%%%%%%%%%%
\section{Introduction}{\label{sec1}}
Tracer dilution curves have emerged as an important tool in visualizing and understanding the dynamics of blood flow and circulation within the body. In this study, we explore the derivation, characteristics, relationships, and applications of tracer dilution curves and perfusion data.
The foundation of this study is based on the works of Davenport \cite{davenport1983derivation} and  MT Madsen \cite{madsen1992simplified}, whose work focused on the derivation of the gamma variate curve and the least squares method of estimating unknown parameters from given gamma variate, respectively.

MT Madsen's \cite{madsen1992simplified} work, builds upon and addresses the limitations identified in Davenport's \cite{davenport1983derivation} previous study which appears to be sensitive to changes in the $\alpha$ and $\beta$ parameters, resulting in unpredictable alterations in the behavior of tracer dilution curves under varying parameter values. One of the key focuses of this study is to observe the changes in the tracer dilution curves under different parameters and establish mathematical principles to recover the parameters and any useful information from the tracer dilution curve given a set of input functions.

In section \ref{sec2}, we discuss the derivative of the gamma variate function and its relationship to tracer dilution curves along with their underlying assumptions. This section also outlines the process of the least squares method and the estimations of the gamma variate parameters for the reconstruction of the true underlining gamma curve and presents the procedures for estimating the parameters of a noisy gamma variate curve. 

In section \ref{sec3} and \ref{sec4}, a comparison of the least squares method and the weighted least squares method for estimating the gamma variate curve is presented together with a method of weighting noisy perfusion data to curb the over-diffusion of signals due to the logarithm transformation from the least squares method.

Section \ref{sec5} and \ref{sec6} outlines the conclusions and focus on future work for better estimation of MRI parameters to enhance the study of perfusion analysis.

%%%%%%%%%%%%%%%%%%%%%%%%%%%%%%%%%%%%%%%%%%%%%%%%%%%%%%%%%%%%%%%%%%%%%%%%%%%%%%%%
\section{Methodology}{\label{sec2}}
\subsection{The Derivation of The Gamma-Variate Relationship for Tracer Dilution Curves}

The derivation of the distribution of the tracer dilution process, its relationship, and its properties as well as its application in organ blood flow and circulation was presented by Robert Davenport in 1983. Davenport \cite{davenport1983derivation} described the gamma variate as a mathematical function that can be used to describe tracer dilution curves. The gamma variate is expressed as 

\begin{align}
    f(t)  = \frac{t^{\alpha}e^{\frac{- t}{\beta}}}{\beta^{(\alpha + 1)}\Gamma(\alpha + 1)},\ \ \alpha > - 1, \beta > 0
    \label{eq1}
\end{align}

where $\alpha$ and $\beta$ are parameters, and $\Gamma(\alpha + 1)$ is a gamma function described as
 \begin{align}
     \Gamma\left(\alpha+1\right)=\int_{0}^{\infty}{x^\alpha e^{-x}dx}
     \label{eq2}
 \end{align}

\subsection{Assumptions}
The blood vessel is made up of a series of \( n \) unknown chambers at equal volumes \( V \), flowing at \( Q \) units at time \( t \). The amount of tracer in a chamber after a change in time (\( D_i(t+\Delta t) \)) equals the amount in the chamber at time \( t \) (\( D_i(t) \)), plus the amount that has flowed from the previous chamber minus the amount that has flowed into the next chamber. Mathematically, we have
\begin{align}
    D_{i}(t+\Delta t) = D_{i}(t) + QV(D_{i-1}(t)\Delta t - D_{i}(t)\Delta t)
    \label{eq3}
\end{align}
Dividing through (\ref{eq3}) by \( V \) and rearranging gives us.

\begin{align}
    C_{i}(t+\Delta t) - C_{i}(t)\Delta t = QV(C_{i-1}(t) - C_{i}(t))
    \label{eq4}
\end{align}
Taking the limits of (\ref{eq4}) as $\Delta t$ approaches infinity gives us. 
\begin{align}
    \frac{d}{dt}C_i(t) = \frac{Q}{V}C_{i-1}(t) - \frac{Q}{V}C_i(t)
\label{eq5}
\end{align}
The assumption that the concentration of tracer in the blood flowing into the first chamber is zero conditions the amount of tracer in the first chamber as
\begin{align}
    \frac{d}{dt}C_1\left(t\right)=-\frac{Q}{V}C_1\left(t\right)
    \label{eq6}
\end{align}
\begin{align}
    C_1\left(t\right)=C_0e^{-Qt/V}
    \label{eq7}
\end{align}
Where $C_0$ is an arbitrary constant. For any other chambers the general pattern is given as 
\begin{align}
    C_n\left(t\right)=\frac{C_0}{\left(n-1\right)!}\left(\frac{Q}{V}t\right)^{n-1}e^{-Qt/V}
    \label{eq8}
\end{align}
It is interesting to note that the gamma variate is in the continuous form, however, the real-life scenario of the blood flow is typically discrete. Davenport \cite{davenport1983derivation} defined the continuous parameters $\alpha$ and $\beta$ to be
\begin{align}
    \alpha\ \ =n\ -1
    \label{eq9}
\end{align}

\begin{align}
    \beta=\ V/Q
    \label{eq10}
\end{align}
This makes it suitable for $\alpha$ and $\beta$ to take on real positive values without any setbacks. The drawback here is that the values of $n$ and $V$ are to be chosen and not determined by nature.

\begin{figure}[h]
\centering
\includegraphics[width=8.5cm, height= 6.3cm]{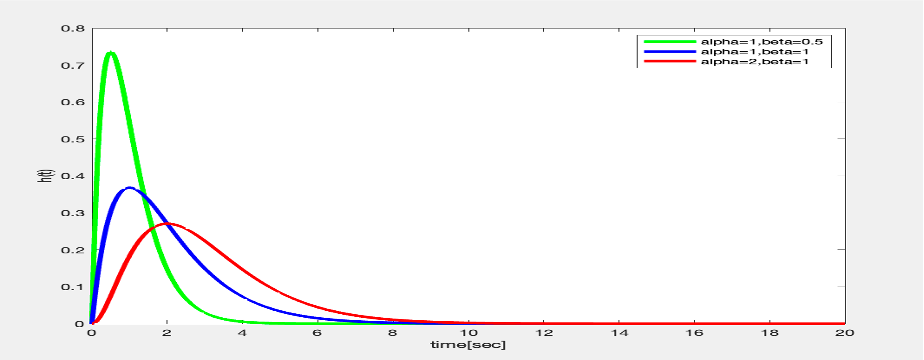}
\caption{A Gamma variates graph for different alpha and beta values}
\label{fig1}
\end{figure}

\begin{figure}[h]
\centering
\includegraphics[width=8.5cm, height= 6.3cm]{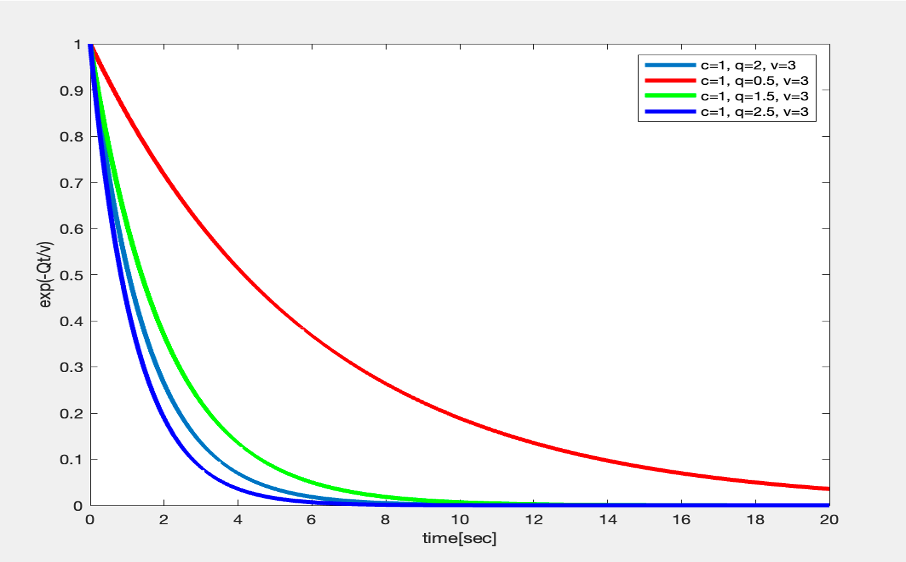}
\caption{Exponential plot of tracer in chamber one}
\label{fig2}
\end{figure}

\begin{figure}[h]
\centering
\includegraphics[width=8.5cm, height= 6.7cm]{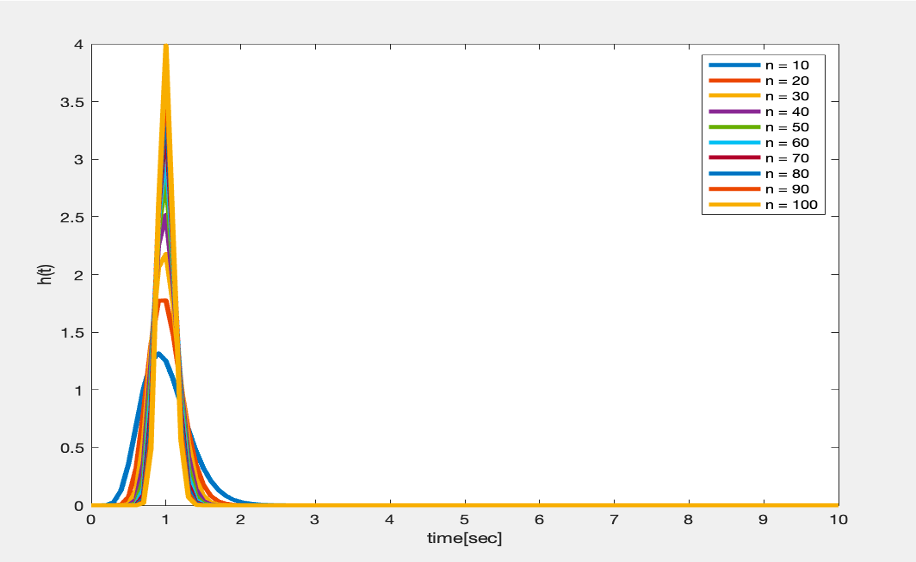}
\caption{The Gamma variate function for increasing $n$ values depicting a Dirac delta function as $n$-approaches infinity}
\label{fig3}
\end{figure}

\begin{figure}[h]
\centering
\includegraphics[width=8.5cm, height= 6.7cm]{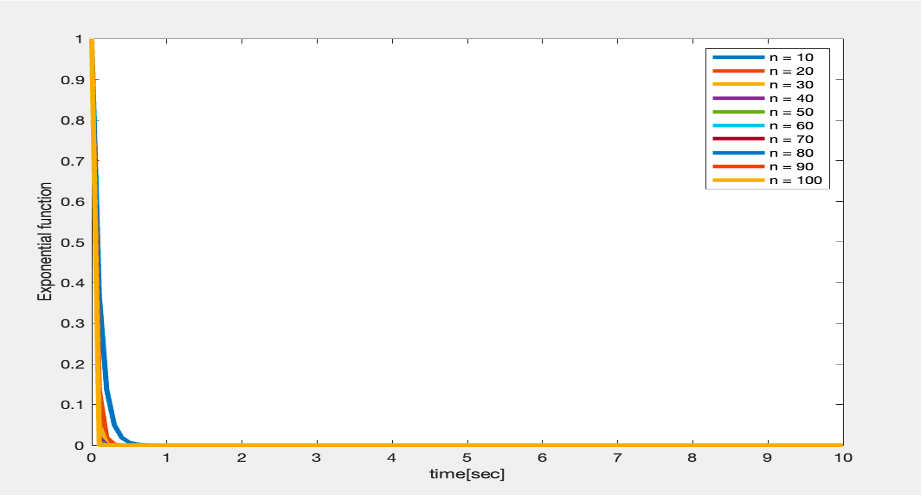}
\caption{The exponential function graph for increasing values of $n$}
\label{fig4}
\end{figure}

%\newpage
In the solution sets for the gamma variate for increasing values of $n$ as shown in Fig. \ref{fig3}, the gamma variate sharpen to a spike depicting a Dirac delta function while the exponential function in Fig. \ref{fig4} approximates an $L$ shape. Suppose that we have a blood vessel with $n$-segments, the transition from one segment to another follows the gamma variate function irrespective of the number of seconds you choose. This conclusion assumes that each segment has equal volume, hence, setting the initial volume to be a constant would require the solution of the Ordinary Differential Equation (ODE) which is a possibility not covered by the in this study. For larger $n$ values the results are not the same (not consistent) as they kept changing due to the variation of the input boundary conditions within each segment. That is, since the volume flowing from segment $i$ to segment $i+1$ keeps changing, how fast or slow the decay is dependent on the size of the change in volume.

\subsection{A simplified formulation of the Gamma Variate Function}

In the Davenport \cite{davenport1983derivation} study, an input signal $x\left(t\right)$ can be said to convolve with an impulse function $h(t)$ over the limits $0$ to $t$. That is

\begin{align}
    y\left(t\right)=\int_{0}^{t}x\left(t\right)h\left(t-\tau\right)d\tau
    \label{eq11}
\end{align}

\begin{align}
    y(t)\ =\ x(t)\otimes h(t)
    \label{eq12}
\end{align}
for periodic functions. The general gamma variate is of the form

\begin{align}
    h(t) = A(t - t_0)^\alpha \exp\left(-\frac{t - t_0}{\beta}\right)
\label{eq13}
\end{align}

Since time $(t)$ is an increasing function its multiplication with the exponential function produces a gamma variate function that rises and falls. The curve is a function of $A$,$\ t_0$, and $\beta$ that would have to be estimated. Notice that the gamma variate function is a continuous function but the circumstances underlying real data points are discretized and characterized with noise. The goal is to estimate the values of $A$,$\ {\ t}_0$, and $\beta$ that best describe the curve by using the least squares method as proposed by Madsen \cite{madsen1992simplified}  by modifying the data points to be linear. The objective for such a transformation is to estimate $\alpha$ as the slope of the linear regression line.
Suppose $\ t_0=0$ because having a non-zero $\ t_0$ shifts the curve to the right,  $t_{max}$ is the value of $t$ for which the gamma variate attains a maximum, and $y_{max}$ is the maximum value of the gamma variate.

\begin{align}
    t_{max}\ =\ \alpha\ast\beta
    \label{eq14}
\end{align}

\begin{align}
    \beta=\ \frac{t_{max}\ }{\ \alpha}
    \label{eq15}
\end{align}

\begin{align}
    y\left(t\right)=A{\ t}^\alpha\ exp(-\alpha t/t_{max}\ )
    \label{eq16}
\end{align}

For $y_{max}$
\begin{align}
    y_{max}\left(t\right)=y\left(t_{max}\right)=A{(\ t_{max})}^\alpha\ exp(-\alpha)
    \label{eq17}
\end{align}
Solving for A from (\ref{eq17}) gives:
\begin{align}
    A\ =\ y_{max}t_{max}^{-\alpha}e^\alpha
    \label{eq18}
\end{align}

substituting $A$ into eq. (\ref{eq16})  yield 

\begin{align}
    y\left(t\right)=y_{max}t_{max}^{-\alpha}e^\alpha({\ t}^\alpha\ exp(-\alpha t/t_{max}\ ))
    \label{eq19}
\end{align}
by setting
\begin{align}
    t^\prime\ =\ \frac{t}{t_{max}}
    \label{eq20}
\end{align}
 we get

 \begin{align}
     y(t^\prime)\ =y_{max}t_{max}^{-\alpha}exp(\alpha)t_{max}^\alpha\ t^{\prime\alpha}\ exp(-\alpha t^{\prime\ })
     \label{eq21}
 \end{align}
taking the natural logarithm yields 

\begin{align}
    ln{\left(y\left(t^\prime\right)\right)}= ln{\left(y_{max}\right)}+(\alpha(1+ ln (t^\prime)-t\prime))
    \label{eq22}
\end{align}

which can be expressed in the least squares matrix form as

\begin{align}
    y=\left[\begin{matrix}y_1\\\vdots\\\vdots\\\end{matrix}\right],\ \ C=\ \left[\begin{matrix}1&(1+\ln (t_1^\prime)-t_1^\prime)\\1&(1+\ln (t_2^\prime)-t_2^\prime)\\\vdots&\vdots\\\end{matrix}\right],\ \ x=\left[\begin{matrix}\ln{\left(y_{max}\right)}\\\alpha\\\end{matrix}\right]
    \label{eq23}
\end{align}

\begin{align}
    \left[\begin{matrix}y_1\\\vdots\\\vdots\\\end{matrix}\right]=\left[\begin{matrix}1&(1+\ln (t_1^\prime)-t_1^\prime)\\1&(1+\ln (t_2^\prime)-t_2^\prime)\\\vdots&\vdots\\\end{matrix}\right]\left[\begin{matrix}\ln{\left(y_{max}\right)}\\\alpha\\\end{matrix}\right]
    \label{eq24}
\end{align}

\begin{align}
    y\ =\ C\cdot x
    \label{eq25}
\end{align}

\begin{align}
    x\ =\ \left(C^TC\right)^{-1}C^Ty
    \label{eq26}
\end{align}

to solve for $y_{max}$ and $\alpha$.

\subsection{Estimating $t_{max}$ and $y_{max}$ from a graph}
Estimating the values of $t_{max}$ and $y_{max}$ enables us to estimate the values of $t^\prime$,  and subsequently  $\widetilde{t}$, before adopting the transformation via eq. (\ref{eq22}) to generate the straight line which will be used to estimate the parameters of the gamma variate $\widetilde{t}\ $ is defined as.

\begin{align}
    \widetilde{t}=(1+\ ln (t^\prime)-t\prime)
    \label{eq27}
\end{align}

The mapping in eq. (\ref{eq27}) as shown in Fig. \ref{fig8}, separates the time domain into two leaves, one from $0$ to $t_{max}$ and the other from $t_{max}$ to the end time of $t$. Accordingly, there will be different values of  $t^\prime$ corresponding to the same $\widetilde{t}$ values, thus we do not see a clearer picture of how this correlate to $y\left(t^\prime\right)$.  Since $\widetilde{t}$ attains its maximum at $t^\prime=1$, it is only monotonic for $t^\prime < 1 $ or $t^\prime>1$. Setting $t_0$ to a value greater than $0$ results in a kink in the log plots which may require estimating for $t<t_{max}$ and $t>t_{max}$.

In the presence of noise, the log plots as indicated in Fig. \ref{fig10} exhibit an interesting trend in the estimation of the slope and intercept. From the plots, the slopes for each noise level fall exactly on each other whereas the offset (intercept) changes at every level of noise. Thus, at high level of noise, the reconstruction of $y_{max}$ from the noisy data may not be a good estimate because of the influence the level of noise has. However, there is robustness in the nature of the mappings in eq. (\ref{eq27}) since the different parameter values do no tend to change the magnitude of $\widetilde{t}$ and the location of $t^\prime$.

\begin{figure}[H]
\centering
\includegraphics[width=8.5cm, height= 5cm]{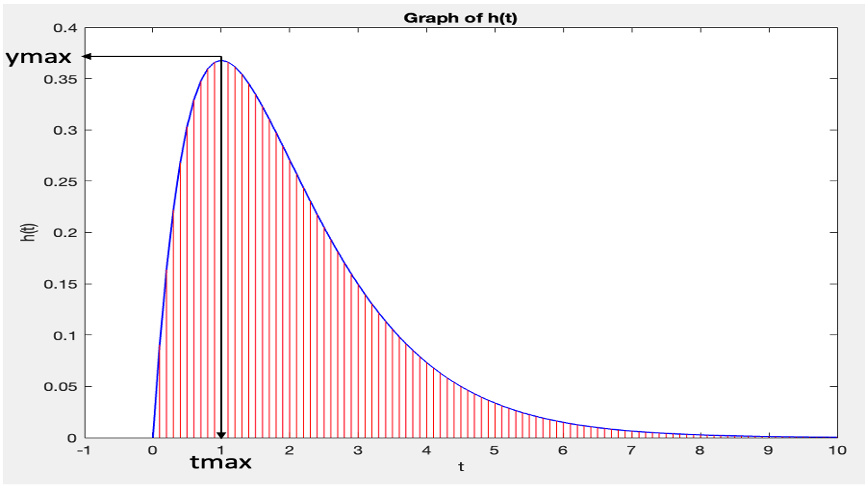}
\caption{A discretized Gamma variate curve showing the location of $t_{max}$ and the magnitude of $h(t))$}
\label{fig5}
\end{figure}

\begin{figure}[h]
\centering
\includegraphics[width=8.5cm, height= 5cm]{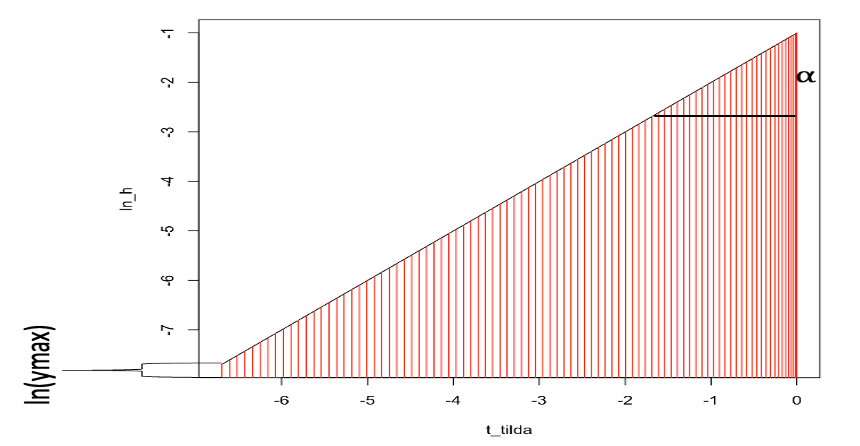}
\caption{log of the discretized data points using the least squares method by Madsen. $ln(y_{max} )$ being the intercept and $\alpha$ is the slope}
\label{fig6}
\end{figure}

\begin{figure}[H]
\centering
\includegraphics[width=8.5cm, height= 5cm]{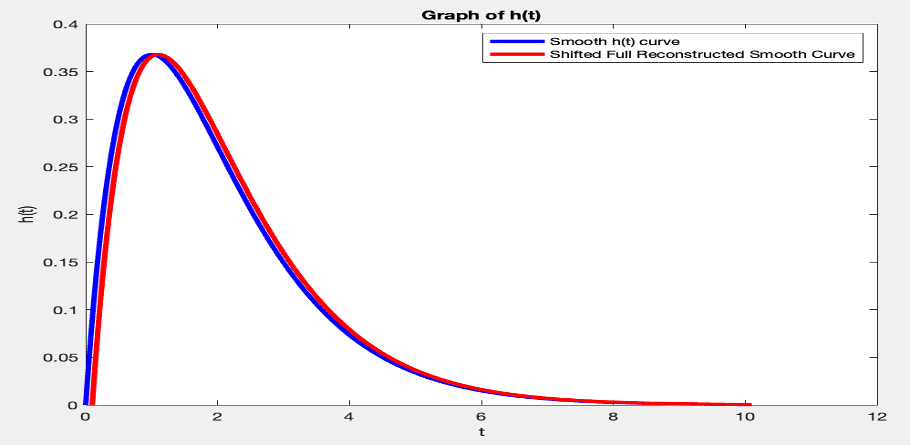}
\caption{Plot of an original gamma variate and estimated curve from the least square method showing the two curves are similar without noise addition. A slight horizontal offset was used to distinguish between the curves for better visibility}
\label{fig7}
\end{figure}

\begin{figure}[h]
\centering
\includegraphics[width=8.5cm, height= 6cm]{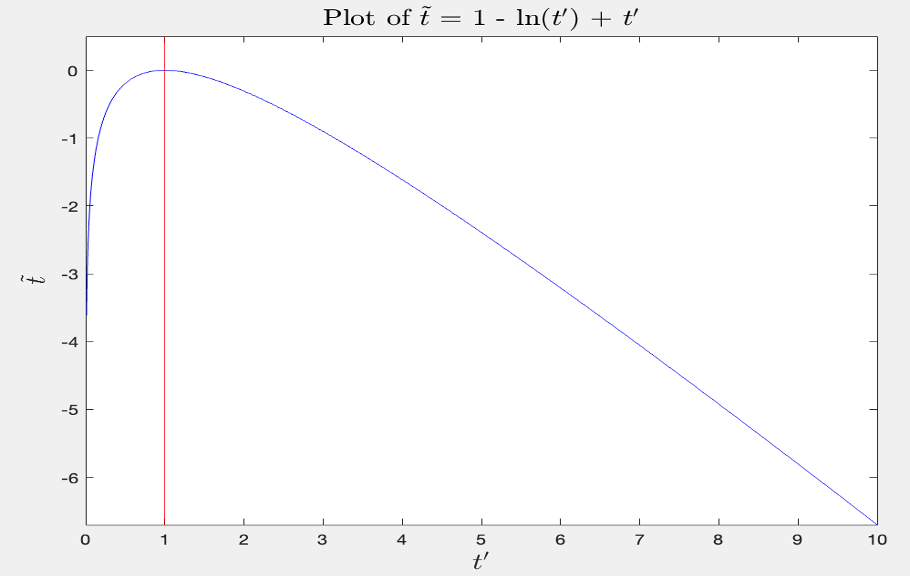}
\caption{Plot of $\widetilde{t}$ against $t^\prime$ showing a monotonic increase $\widetilde{t}$ at  $t^\prime>1$}
\label{fig8}
\end{figure}

\begin{figure}[H]
\centering
\includegraphics[width=8.5cm, height= 6cm]{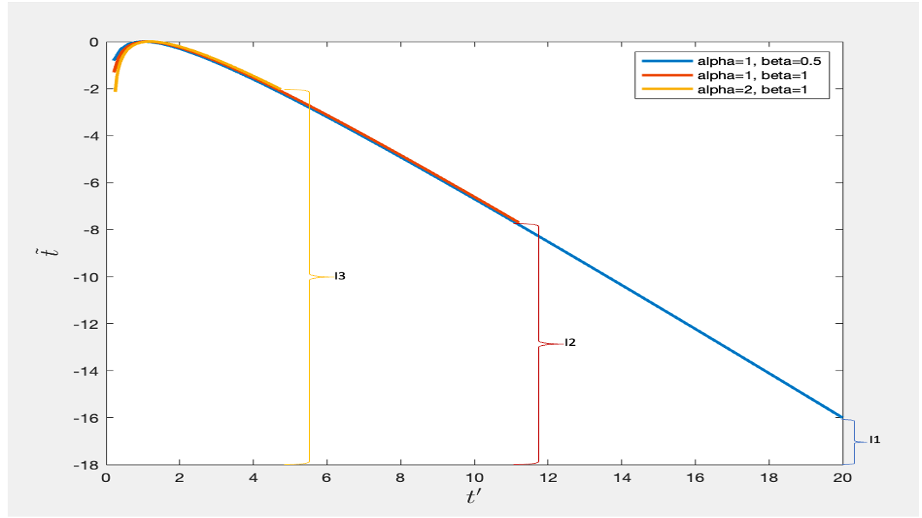}
\caption{Logarithm plot of eq. (\ref{eq27}) for different $\alpha$ and $\beta$ values at a noise level of 0.01 showing the same slope but increasing intercepts (I1, I2, I3). Plots have been slightly offset for good visibility.}
\label{fig9}
\end{figure}

\begin{figure}[h]
\centering
\includegraphics[width=8.5cm, height= 6cm]{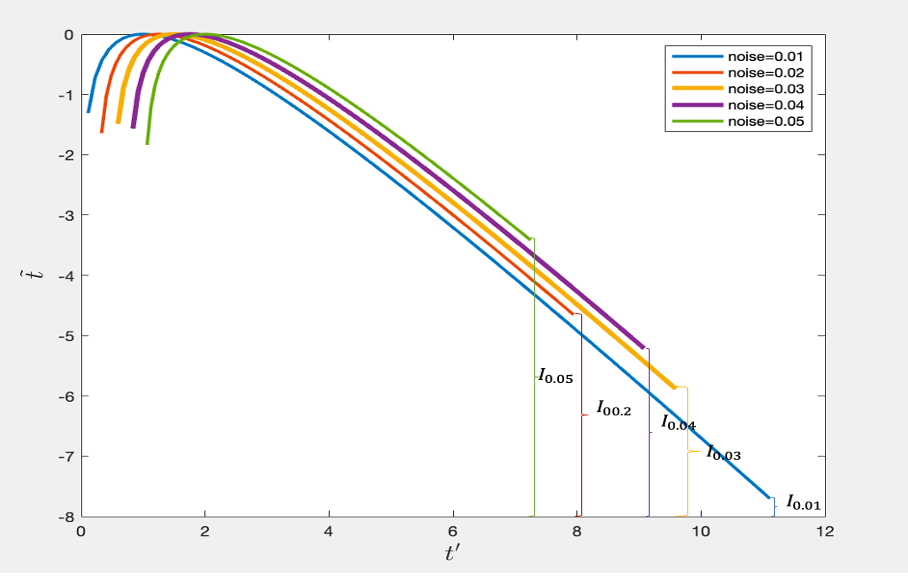}
\caption{Logarithm plot of eq. (27)  for different noise levels with  $I_{0.01}$, $I_{0.02}$, $I_{0.03}$, $I_{0.04}$, and $I_{0.05}$ being the intercepts for the noise levels 0.01 through to 0.05. Plots have been slightly offset for good visibility}
\label{fig10}
\end{figure}

%\newpage
\section{Results}{\label{sec3}}
\subsection{Estimating the Parameters of a noisy Gamma variate curve with the Madsen procedure}

\begin{figure}[H]
\centering
\includegraphics[width=8.5cm, height= 6cm]{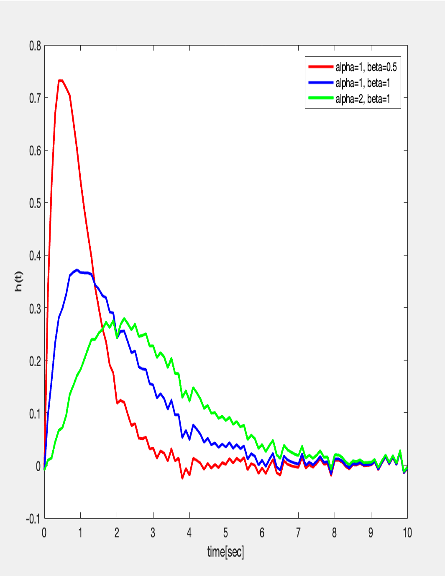}
\caption{The gamma variate curve with noise for varying $\alpha$ and $\beta$ values}
\label{fig11}
\end{figure}

Suppose that a gamma variate curve is made up of noise randomly sampled from the normal distribution with mean $0$ and standard deviation $0.01$ over varying $\alpha$ and $\beta$ values as shown in Fig. \ref{fig11}, The Madsen procedure provides a robust way to estimate parameters, reducing the impact of outliers and noise in the data. By utilizing this approach, the accuracy and reliability of the parameter estimation from the least square methods are improved significantly. The steps involved in estimating the parameters are.

\begin{itemize}
    \item Plotting the noisy data to extract \( t_{\text{max}} \) and \( y_{\text{max}} \)
    \item Estimating \( t^\prime \) and \( \widetilde{t} \)
    \item Solving the least squares method in (23)
\end{itemize}

\begin{figure}[H]
\centering
\includegraphics[width=8.5cm, height= 6.5 cm]{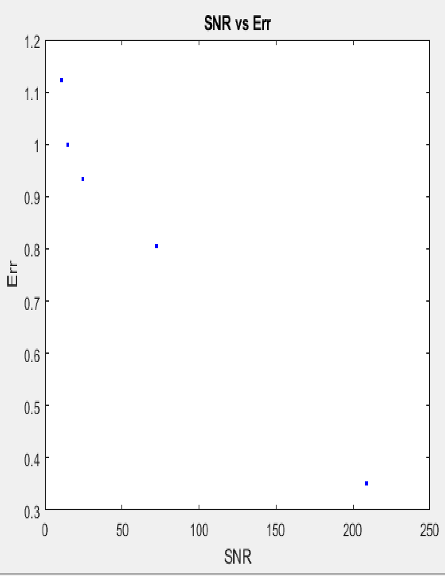}
\caption{SNR vs Goodness of Fit plot}
\label{fig12}
\end{figure}

The goodness of fit (gof) and signal-to-noise ratio (SNR) plot is constructed by plotting the vector norms of the difference in the original data with noise and the reconstructed curves to ascertain the true relationship between the goodness of fit and noise. The SNR is a measurement of the performance of MRI systems and is used to evaluate contrast performance \cite{dietrich2007measurement}. In this study, the SNR is taken from five different noise levels at standard deviations from 0.01 to 0.05. Fig. \ref{fig12} depicts the plot of the goodness of fit vs. the SNR showing a negative relationship between increasing SNR and the error which is the vector of the difference in the original curve and reconstructed curves. The SNR is given by.

\begin{align}
    SNR\ =\ \frac{E({h(t)}^2)}{var(noise)}
    \label{eq28}
\end{align}

where $h(t)$ is the gamma variate function with noise.  The steps to computing the SNR for the case studies are outlined below.

\begin{enumerate}
    \item Construct a gamma variate curve $(h(t))$ with noise $\sim$ $N(0,0.01)$
    \item Use the Madsen procedure to reconstruct the true gamma curve $(h_{new}(t))$
    \item Apply eq. (28) to calculate the SNR
    \item 	Compute the norm of the difference between the original curve and the reconstructed curve $(h_{new}(t)\ -\ h(t))$
    \item Repeat steps 1 to 4 for different added noise levels from 0.02 through to 0.05
    \item Plot step 4 for the various noise levels against step 3
\end{enumerate}

by reconstructing the noisy curve, we can extract meaningful information for analysis and visualize the underlying pattern of the signal. In table \ref{tab1} below, we show the estimated $\beta$ and $\alpha$ values from the noisy signals with the Madsen procedure and \ref{tab2} gives an overview of the SNR and error values for the construction of Fig. \ref{fig12}

\begin{table}[ht]
\centering
\footnotesize
\caption{Estimates of $\beta$ and $\alpha$ values from the Madsen procedure and WLS with 1\% added noise.}
\setlength\tabcolsep{3pt}  % <---
\label{tab1}
  \begin{tabular}{@{} l*{6}{c} @{}}
    \hline
\textbf{Parameter} & \textbf{Original}   &   &  
                &   \textbf{Madsen Estimation}  &     \\ \cline{3-6}

    \textbf{Estimates}    & \textbf{values} & OLS       & Relative & WLS & Relative     \\ 
            &   &        & deviation &  & deviation  \\
    \hline
    $\beta$ & 1.0 & 1.072 & 7.23\% & 0.9999 & 0.01\%   \\
    \hline
    $\alpha$ & 1.0 & 0.9491 & 5.09\% & 1.0174 & 1.74\%   \\
    \hline
    
    $y_{max}$ & 0.3679 & 0.3518 & 4.38\% & 0.3664 & 0.41\% \\
    \hline
  \end{tabular}
\end{table}

\begin{table}[h]
\caption{SNR values with corresponding errors and randomly added noise from the normal distribution}
\label{tab2}
\begin{center}
\begin{tabular}{|c||c||c|}
\hline
Noise & SNR & Norm (Error)\\
\hline
$N(0, 0.01)$ &183.5301  & 0.31192 \\
\hline

$N(0, 0.02)$ &63.0594     & 0.7724 \\
\hline

$N(0, 0.03)$ &24.8804     & 0.8813 \\
\hline

$N(0, 0.04)$ &15.3563     & 1.0240 \\
\hline

$N(0, 0.05)$ &12.3746  & 1.0174 \\
\hline
\end{tabular}
\end{center}
\end{table}

\subsection{The Ordinary Least Squares (OLS) and Weighted Least Squares (WLS) methods}

\onecolumn
\begin{multicols}{2}
Linear regression is a model to explain the linear relationship between an independent variable and a set of one or multiple dependent variables.  This is a useful technique for predicting outcomes based on a set of values. For an accurate estimation of the parameters from the linear regression, the following assumptions must be met or otherwise the variables must be transformed to meet this assumptions.\\
\emph{Assumptions of the OLS:}
\begin{enumerate}
    \item A linear relationship between the independent and dependent variables
    \item Independence of error terms
    \item The independent and dependent variables must be independent of each other
    \item Homoscedasticity: Residuals have constant variance across all levels of the independent variables
    \item Normality: Residuals are normally distributed
\end{enumerate}
The regression model is of the form.  
\begin{align}
    Y = \mathbf{\beta}_0 + \mathbf{\beta}_1 X_1 + \mathbf{\beta}_2 X_2 + \ldots + \mathbf{\beta}_i X_i + \mathbf{\epsilon}_i
    \label{eq29}
\end{align}

Where $Y$ is the dependent variable, ${\beta}_{0}$ is the intercept, ${\beta}_{i}$  are parameters to be estimated from the model, ${\epsilon}_{i}\ $ are the random error terms ${N}$ $({0}$,${\sigma}_{i}^{2})$, and ${X}_{i}$ are the independent variables. The OLS estimation formula for the coefficients is derived by minimizing the sum of squared residuals (SSR) via eq. (23).

\subsection{Weighted Least Squares}
The OLS assumption of constant variance (homoscedasticity) of residuals is not really the case in real-world data since they exhibit heteroscedasticity (non-constant variance).  The WLS addresses this by assigning higher weights to observations with smaller variances and lower weights to observations with larger variances \cite{kutner2005applied}.
The WLS estimation formula accounts for the weighted sum of squared residuals:

\begin{align}
    \beta_{\left(WLS\right)} = \left((X^TW^{-1})^{-1}\right)X^TW^{-1}y
    \label{eq30}
\end{align}

\subsection{Estimating the Weights for WLS}
The variance-covariance matrix of the error terms in eq. (\ref{eq29}) is given as 
\begin{align}
    \sigma_{nxn}^2\ =\ \left(\begin{matrix}\sigma_1^2&\cdots&0\\\vdots&\ddots&\vdots\\0&\cdots&\sigma_n^2\\\end{matrix}\right)
    \label{eq31}
\end{align}

Assuming that the error terms are known, the methods of maximum likelihood can be used to estimate the regression coefficients in eq. (\ref{eq26}) by simply replacing the equal error variance $\sigma^2$ in the likelihood function with the respective variance $\sigma_i^2$ values.  The maximum likelihood function for eq. (\ref{eq26}) is given as
\end{multicols}

\begin{align}
 L\left(\beta_0,\beta_1,\ldots,\beta_k,\sigma^2\right)=\prod_{i=1}^{n}\frac{1}{\sqrt{2\pi\sigma^2}}\exp{\left(-\frac{\left(y_i-\left(\beta_0+\beta_1x_{i1}+\beta_2x_{i2}+\ldots+\beta_kx_{ik}\right)\right)^2}{2\sigma^2}\right)}
    \label{eq32}
\end{align}
%\end{strip}

replacing $\sigma^2$ with $\sigma_i^2$ gives us

%\begin{strip}
    \begin{align}
        L\left(\beta_0,\beta_1,\ldots,\beta_k,\sigma_i^2\right)=\prod_{i=1}^{n}\frac{1}{\sqrt{2\pi\sigma_i^2}}\exp{\left(-\frac{\left(y_i-\left(\beta_0+\beta_1x_{i1}+\beta_2x_{i2}+\ldots+\beta_kx_{ik}\right)\right)^2}{2\sigma_i^2}\right)}
        \label{eq33}
    \end{align}
%\end{strip}

\begin{align}
    \log{L\left(\beta_0,\beta_1,\ldots,\beta_k,\sigma_i^2\right)=\frac{1}{2}\sum_{1=1}^{n}\log{\left(\frac{w_i}{2\pi}\right)-\frac{1}{2}\sum_{i=1}^{n}{w_i\left(y_i-\beta_0-\ \beta_1x_{i1}-\beta_2x_{i2}-\ldots-\beta_kx_{ik}\right)^2}}}
    \label{eq34}
\end{align}

    by denoting the weight as $w_i=\frac{1}{\sigma_i^2}$ we can modify the eq. (\ref{eq31}) as.

\begin{align}
    L(\beta_0,\beta_1,\ldots,\beta_k,\sigma_i^2) = \left[\prod_{i=1}^{n}\left(\frac{w_i}{2\pi}\right)^{\frac{1}{2}}\right]\exp\left[-\frac{1}{2}\sum_{i=1}^{n}w_i\left(y_i-\beta_0-\beta_1x_{i1}-\beta_2x_{i2}-\ldots-\beta_kx_{ik}\right)^2\right]
    \label{35}
\end{align}

taking the log-likelihood transformation of eq. (\ref{eq32}) gives us.

\begin{align}
    L\left(\beta_0,\beta_1,\ldots,\beta_k,\sigma_i^2\right)=\ \left[\prod_{i=1}^{n}\left(\frac{w_i}{2\pi}\right)^{{\frac{1}{2}}}\right]exp\left[-\frac{1}{2}\sum_{i=1}^{n}{w_i\left(y_i-\beta_0-\ \beta_1x_{i1}-\beta_2x_{i2}-\ldots-\beta_kx_{ik}\right)^2}\right]
    \label{eq36}
\end{align}

to maximize the log-likelihood we differentiate eq. (33) with respect to parameters and set to zero to give us

\begin{align}
    \log L(\beta_0,\beta_1,\ldots,\beta_k,\sigma_i^2) = \frac{1}{2}\sum_{i=1}^{n}\log\left(\frac{w_i}{2\pi}\right) - \frac{1}{2}\sum_{i=1}^{n}{w_i\left(y_i-\beta_0-\beta_1x_{i1}-\beta_2x_{i2}-\ldots-\beta_kx_{ik}\right)^2}
    \label{eq37}
\end{align}

\multicols{2}
which are the weighted least squares. Notice that the weighted least squares depict the case of the ordinary least squares with a weight value $w_i$ rather than 1. Since the weights $(w_i)$ is the inverse of the $\sigma_i^2$, the observations with large variance receive less weight than observations with less variance. The weighted least squares can be expressed in matrix form as

\begin{align}
    {W\ }_{nxn}=\ \left(\begin{matrix}w_i&\cdots&0\\\vdots&\ddots&\vdots\\0&\cdots&w_n\\\end{matrix}\right)
    \label{eq38}
\end{align}
so, the regression eq. with the weighted least squares can be expressed as

\begin{align}
    \left(X^TWX\right)b_w\ =\ X^TWY
    \label{eq39}
\end{align}
which is used to estimate the regression coefficients as indicated in eq. (30). The variance-covariance matrix of the weighted least squares parameter estimates is.

\begin{align}
    \sigma^2{b_w}=\left(X^TWX\right)^{-1}
    \label{eq40}
\end{align}
The WLS can be performed using the steps below.
\begin{enumerate}
    \item Run an OLS model with the respective $Y$  and $X$ values.
    \item Run a second OLS model with the absolute values of the residual from the first model as the $Y$ values against the $X$ values.
    \item Weights = $\frac{1}{\sigma_i^2}\ =\ \frac{1}{{fitted\ values}^2}$
    \item Run a final OLS regression with the original values of $Y$ and $X$ with a defined weight as stated in step 3.
\end{enumerate}

\subsection{The Effect of Noise }
\begin{figure}[H]
\centering
\includegraphics[width=8.5cm, height= 6 cm]{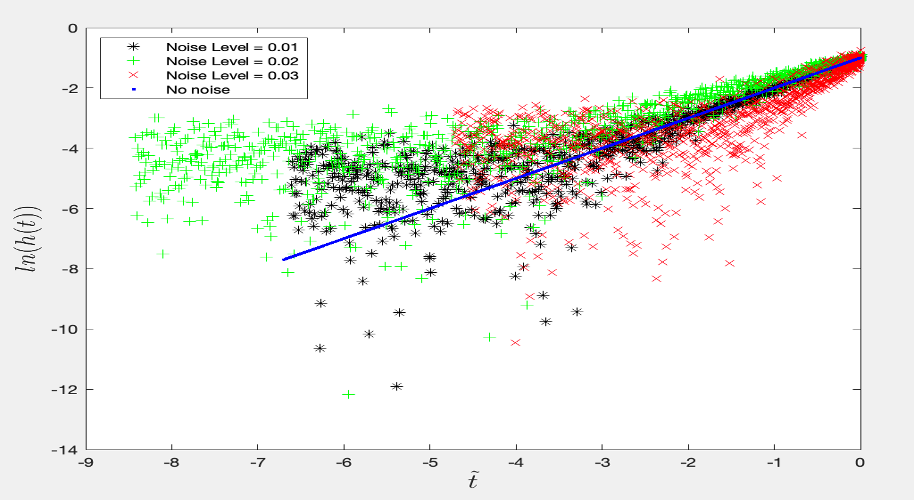}
\caption{Plot of $ln(h)$ for $\alpha$ =1 and $\beta$ = 1 with different noise levels}
\label{fig13}
\end{figure}

\begin{figure}[H]
\centering
\includegraphics[width=8.5cm, height= 6 cm]{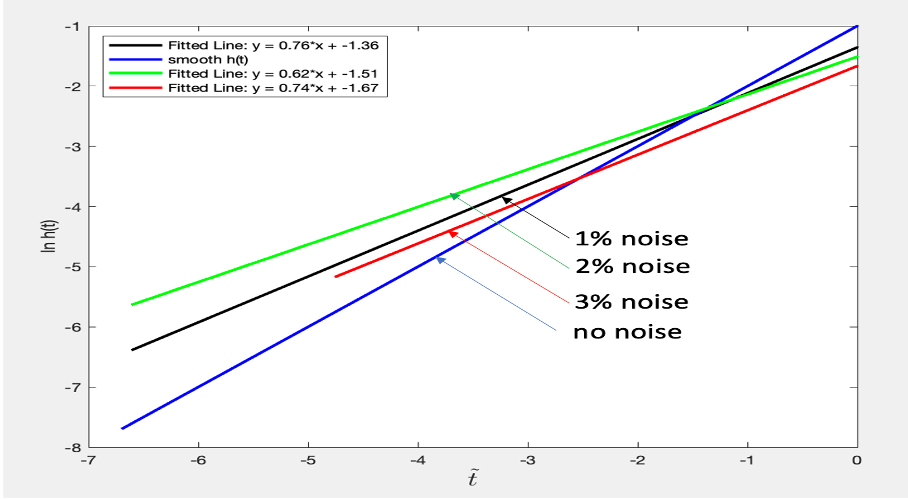}
\caption{A comparison of the fitted regression lines with noise at different levels}
\label{fig14}
\end{figure}

\begin{figure}[H]
\centering
\includegraphics[width=8.5cm, height= 6 cm]{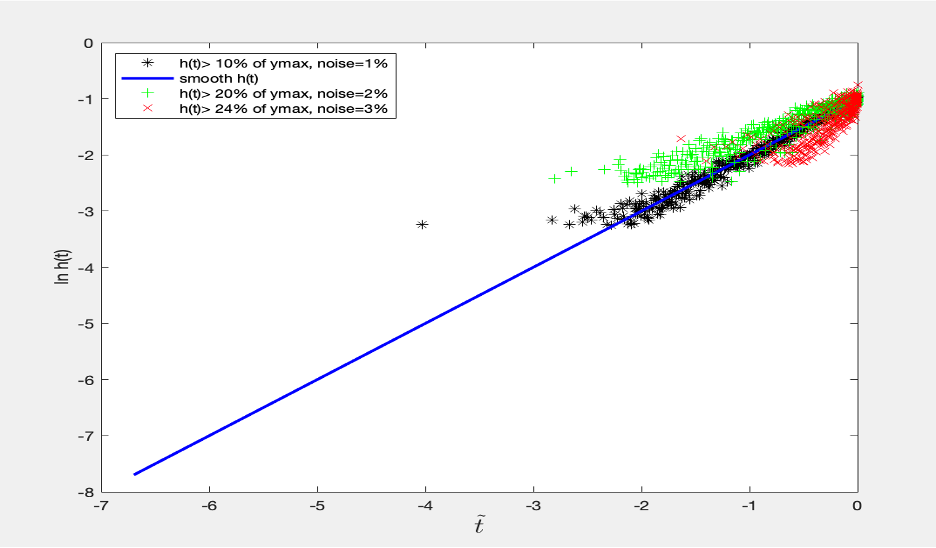}
\caption{A comparison of the fitted regression lines with trimmed $h(t)$ values}
\label{fig15}
\end{figure}

\begin{figure}[H]
\centering
\includegraphics[width=8.5cm, height= 5.8 cm]{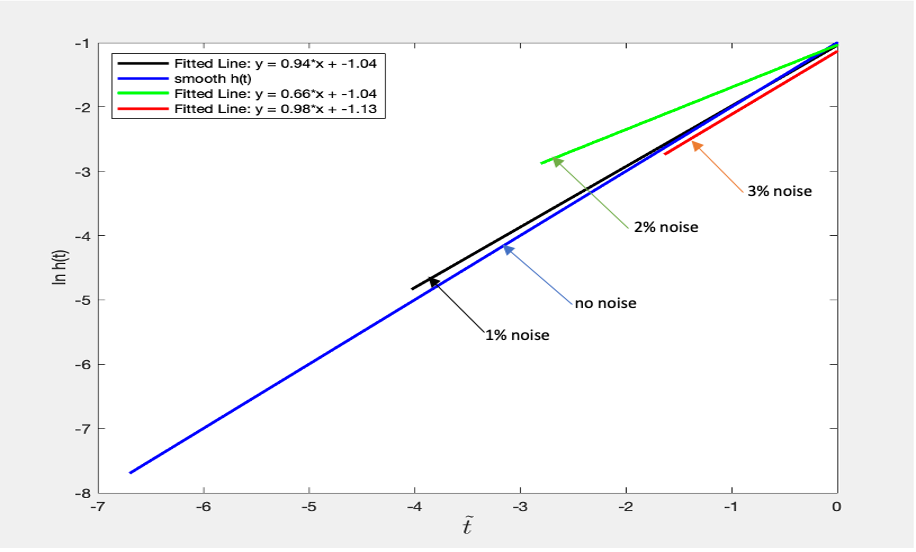}
\caption{A comparison of the fitted regression lines with weights set to zero for all $h(t)$ values less than a stated percentage of $y_{max}$}
\label{fig16}
\end{figure}

In Fig. \ref{fig13}, the graph appears scattered and diffuse, which can influence estimating the slope and offset at different noise levels. The underlying factors for such scatter in the graph are attributed to the exaggeration of the log values of $h(t)$ nearing zero. To address this, all $h(t)$ greater than a percentage of $y_{max}$ would be retained, and those less than the threshold would be assigned a weight value of zero. This procedure curtails the wide diffusion of the graph and brings the estimated slope and offset from such data near the original smooth $h(t)$. Fig. \ref{fig14} shows the fitted regression lines from Fig. \ref{fig13}. As can be seen in Fig. \ref{fig15}, the retained $h(t)$ values appear to be more linear compared to when all $h(t)$ was used and the resulting fitted lines in Fig. \ref{fig16} are much better than those in Fig. \ref{fig13}. Though the weights of zero to the set of $h(t)$ which do not meet the stated condition appears to do a good job in bringing the noisy data close to the original smooth data, there is no definite method or mathematical principle to inform what percentage of $y_{max}$ should be used for such an experiment hence, for this study we tried different threshold values to arrive at the best fit, i.e., 10\% of $y_{max}$  for 1\% noise, 20\% of $y_{max}$  for 2\% noise, and 24\% of $y_{max}$  for 3\% were retained for the plots in Fig. \ref{fig15} and Fig. \ref{fig16}. 

\subsection{Estimation of $y_{max}$, $\alpha$, and  $\beta$}
The process that best estimates $y_{max}$ is the weighted least squares method after all $h(t)$ values less than the stated percentage conditions are set to zero. The intercept from the weighted least squares represents $ln(y_{max})$ (i.e., $e^{intercept}\ =\ y_{max})$ whereas the slope $(\alpha)$  is used to calculate $\beta$ using eq. (\ref{eq15}). The overall goal is to estimate $y_{max}$ to be as close as possible to the maximum $h(t)$ value from the original curve.

\subsection{Assessment of Estimates against Ground Truth}
\begin{figure}[H]
\centering
\includegraphics[width=8.5cm, height= 7cm]{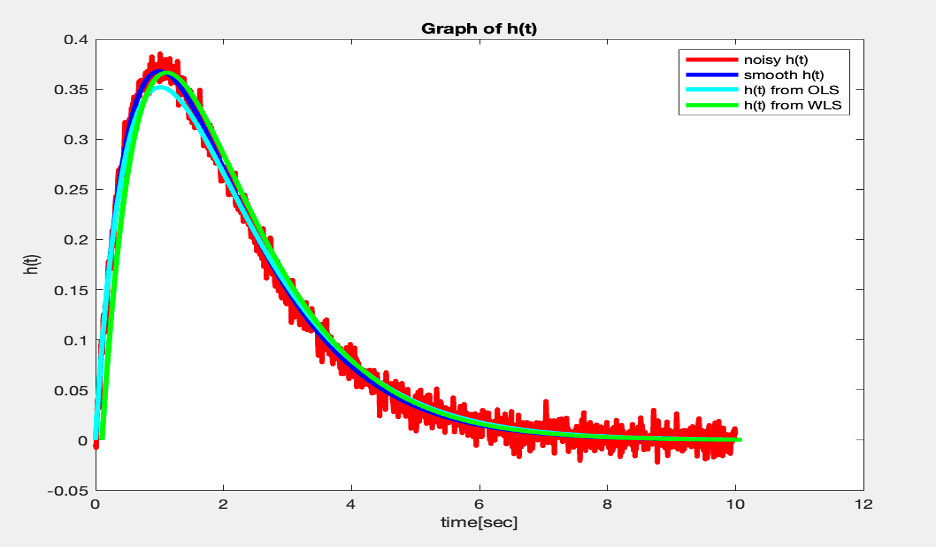}
\caption{Plot of the original curve and reconstructed curve using the OLS and WLS methods. Only $h(t)$ values greater than 10\% of $y_{max}$ were retained for the reconstructions. For better viewing, the reconstructed $h(t)$ from the WLS is slightly shifted}
\label{fig17}
\end{figure}

\begin{figure}[H]
\centering
\includegraphics[width=8.5cm, height= 7cm]{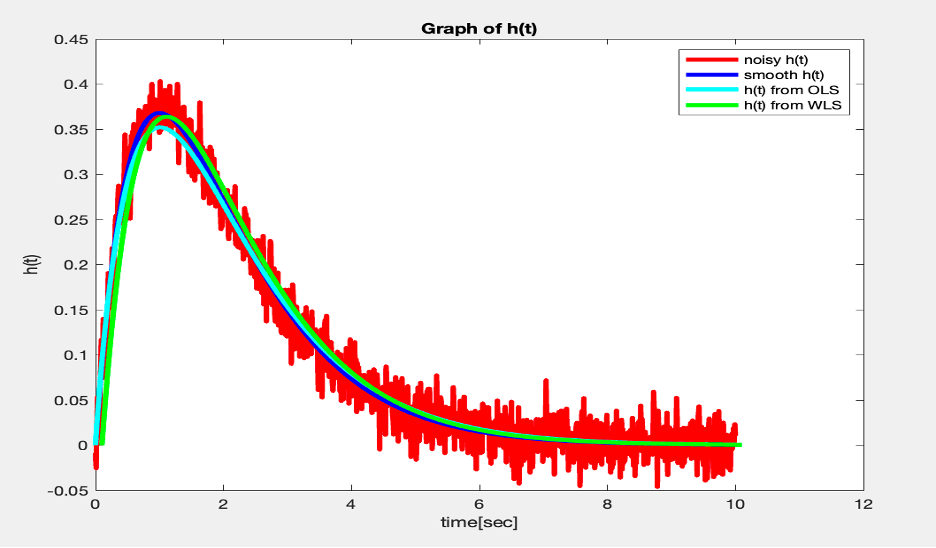}
\caption{Plot of the original curve and reconstructed curve using the OLS and WLS methods. Only $h(t)$ values greater than 20\% of $y_{max}$ were retained for the reconstructions. For better viewing, the reconstructed $h(t)$ from the WLS is shifted slightly}
\label{fig18}
\end{figure}

\begin{figure}[H]
\centering
\includegraphics[width=8.5cm, height= 7cm]{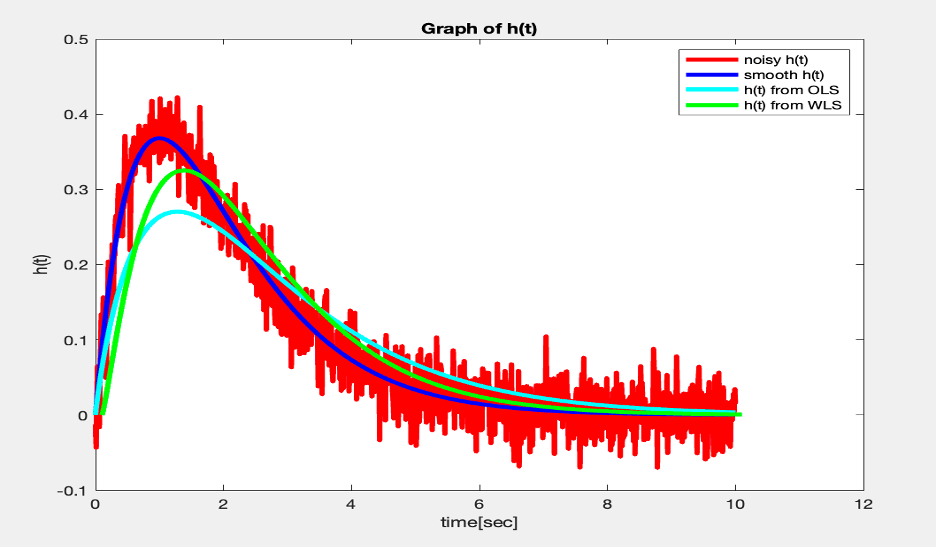}
\caption{Plot of the original curve and reconstructed curve using the OLS and WLS methods. Only $h(t)$ values greater than 24\% of $y_{max}$ were retained for the reconstructions. For better viewing, the reconstructed $h(t)$ from the WLS is shifted slightly}
\label{fig19}
\end{figure}

Fig. \ref{fig17}, Fig. \ref{fig18}, and Fig. \ref{fig19} show the reconstructed curves from the OLS and WLS methods with varying noise levels. In Fig. \ref{fig17} the estimated curve at a noise level of 1\% from the OLS produced a $y_{max}$ value of 0.3518 against the true $y_{max}$ value of 0.3679 whiles the estimated curve from the WLS produced a $y_{max}$ value of 0.3664. These two estimates represent a relative deviation of 4.38\% and 0.41\% from the true $y_{max}$ value for the OLS and WLS methods respectively implying that the WLS method provides a better fit. Similarly, at 2\% and 3\% noise levels, the WLS estimates were much better than the OLS estimates of $y_{max}$, however, the relative deviations are higher compared to when the noise level is 1\% which suggests that the larger the noise levels the less accurate $y_{max}$ value is estimated.  Table \ref{tab3} and Table \ref{tab4} contain the estimates from the OLS and WLS at 2\% and 3\% noise respectively along with their relative deviations.

\begin{table}[H]
\centering
\footnotesize
\caption{Estimates of $\beta$ and $\alpha$ values from the OLS and WLS with 2\% added noise}
\setlength\tabcolsep{3pt}  % <---
\label{tab3}
  \begin{tabular}{@{} l*{6}{c} @{}}
    \hline
\textbf{Parameter} & \textbf{Original}   &   &  
                &   \textbf{Estimations}  &     \\ \cline{3-6}

    \textbf{Estimates}    & \textbf{values} & OLS       & Relative & WLS & Relative     \\ 
            &   &        & deviation &  & deviation  \\
    \hline
    $\beta$ & 1.0 & 1.0715 & 7.15\% & 1.0123 & 1.23\%   \\
    \hline
    $\alpha$ & 1.0 & 0.9426 & 5.74\% & 0.9977 & 0.23\%   \\
    \hline
    
    $y_{max}$ & 0.3679 & 0.3520 & 4.32\% & 0.3638 & 1.11\% \\
    \hline
  \end{tabular}
\end{table}

\begin{table}[H]
\centering
\footnotesize
\caption{Estimates of $\beta$ and $\alpha$ values from the OLS and WLS with 3\% added noise}
\setlength\tabcolsep{3pt}  % <---
\label{tab4}
  \begin{tabular}{@{} l*{6}{c} @{}}
    \hline
\textbf{Parameter} & \textbf{Original}   &   &  
                &   \textbf{Estimations}  &     \\ \cline{3-6}

    \textbf{Estimates}    & \textbf{values} & OLS       & Relative & WLS & Relative     \\ 
            &   &        & deviation &  & deviation  \\
    \hline
    $\beta$ & 1.0 & 1.4230 & 42.3\% & 1.06 & 3.06\%   \\
    \hline
    $\alpha$ & 1.0 & 0.8995 & 10.05\% & 1.2419 & 24.19\%   \\
    \hline
    
    $y_{max}$ & 0.2703 & 0.3520 & 26.53\% & 0.3252 & 11.61\% \\
    \hline
  \end{tabular}
\end{table}

%%%%%%%%%%%%%%%%%%%%%%%%%%%%%%%%%%%%%%%%%%%%%%%%%%%%%%%%%%%%%%%%%%%
%%%%%%%%%%%%%%%%%%%%%%%%%%%%%%%%%%%%%%%%%%%%%%%%%%%%%%%%%%%%%%%%%%%
%%%%%%%%%%%%%%%%%%%%%%%%%%%%%%%%%%%%%%%%%%%%%%%%%%%%%%%%%%%%%%%%%%%
%%%%%%%%%%%%%%%%%%%%%%%%%%%%%%%%%%%%%%%%%%%%%%%%%%%%%%%%%%%%%%%%%%%
%%%%%%%%%%%%%%%%%%%%%%%%%%%%%%%%%%%%%%%%%%%%%%%%%%%%%%%%%%%%%%%%%%%
%%%%%%%%%%%%%%%%%%%%%%%%%%%%%%%%%%%%%%%%%%%%%%%%%%%%%%%%%%%%%%%%%%%

\section{Discussion}{\label{sec4}}

The dynamic change of a tracer converted in the blood through a cylindrical vessel was shown to follow the shape of Gamma variate functions. However, in Davenport’s derivation \cite{davenport1983derivation}, tracer infusion enters as a boundary condition into an initial feeding cell of finite volume. This assumption renders predicted outlet concentration curves dependent on the boundary assumption; thus, Gamma variate functions are not strictly the correct analytical tracer outlet response. In practice, the gamma variate function renders realist concentration curves but does not constitute a practical numerical method for solving tracer convection through a network analytically.

 Madsen \cite{madsen1992simplified} discusses how to fit a Gamma variate function ({with three DOF, $t_0$, $t_{\text{max}}$, $\alpha$, and $\beta$) from experimental tracer dilution curves. Setting $t_0$ and $t_{\text{max}}$ from the data, and using several coordinate transformations, the remaining unknowns, $\alpha$ and $\beta$, can be obtained by linear least square parameter estimation. Our simulation experiments showed that in the presence of noise, the noisy input data need to be weighed so that the values near $t_{\text{max}}$ have the strongest impact, while values far away (with intensities close to zero) need to be de-emphasized or just omitted (zero weight). Without weighting, logarithm errors around small intensity values throw the method off. The process of estimating parameters from the Madsen process is rigorous and robust. We did not establish a definite procedure to choose $t_0$ so it is difficult to ascertain exactly where the curve begins, although we have a reference to do this by the energy method \cite{markalous2008detection}.

\section{Conclusion}{\label{sec5}}
Overall, the Madsen procedure provides a more rigorous and robust way to estimate MRI parameters as it depicts real-world data scenarios by the inclusion of noise and eliminates the issues of fluctuating the magnitude of the tracer dilution curve encompassed by varying parameters, however, the estimation of $y_{\text{max}}$ is better reconstructed using the weighted least squares method after the Madsen process.

\section{Future work}{\label{sec6}}
The focus of this study is to explore the various parameter estimation methods in tracer dilutions curves in medical image processing. Since these methods explored are estimates, we strive to minimize the error associated with each estimate as closely as possible to depict the distribution of the real data we seek to estimate. Hence, other estimation methods would be explored and compared to the methods discussed in this report to arrive at the one with minimal error and much more accuracy.

Overall, the focus of future work should be the accuracy, applicability, and development of a more rigorous way to estimate MRI data parameters to enhance the study of tracer dilution and blood flow.

\section*{Acknowledgment}
I would like to express my sincere gratitude to Dr. Andreas Linninger, the director of the LPPD for the opportunity granted me to undertake an internship at his research lab. Dr. Linninger’s guidance, mentoring, and great supervision throughout the period of my internship has been a great turning point in my academic journey and his great leadership and insight have offered me the opportunity to learn the significant roles statistics play in perfusion analysis, biomedical engineering, and neuroscience in general. 

I would also like to express my appreciation to Thomas Ventimiglia for his invaluable support and guidance throughout the period of my internship. His unwavering commitment to my learning and development has been instrumental in enhancing my understanding of complex concepts and tasks.


%%%%%%%%%%%%%%%%%%%%%%%%%%%%%%%%%%%%%%%%%%%%%%%%%%%%%%%%%%%%%%%%%%%%%%%%%%%%%%%%
%\nocite*
\bibliographystyle{IEEEtran}  
\bibliography{main}

\end{document}